\documentclass[conference]{IEEEtran}
\IEEEoverridecommandlockouts
\usepackage{cite}
\usepackage{amsmath,amssymb,amsfonts}
\usepackage{algorithm}
\usepackage{algorithmic}
\usepackage{graphicx}
\usepackage{caption}
\usepackage{float}
\usepackage{afterpage}
\usepackage{multirow}
\usepackage{graphicx}
\usepackage{subcaption}
\usepackage{textcomp}
\usepackage{colortbl}
\usepackage[table,xcdraw]{xcolor}
\usepackage{xcolor}
\def\BibTeX{{\rm B\kern-.05em{\sc i\kern-.025em b}\kern-.08em
    T\kern-.1667em\lower.7ex\hbox{E}\kern-.125emX}}
\begin{document}

\title{Towards a Proactive Autoscaling Framework for Data Stream Processing at the Edge using GRU and Transfer Learning\\
}

\author{\IEEEauthorblockN{Eugene Armah}
\IEEEauthorblockA{\textit{Computer Science Department} \\
\textit{Kwame Nkrumah University of Science and Technology}\\
\textit{Kumasi, Ghana}\\
Email: earmah16@st.knust.edu.gh}
\and
\IEEEauthorblockN{Linda Amoako Banning}
\IEEEauthorblockA{\textit{Computer Science Department} \\
\textit{Kwame Nkrumah University of Science and Technology}\\
\textit{Kumasi, Ghana}\\
Email: labanning@knust.edu.gh \\
}

}

\maketitle

\begin{abstract}
Processing data at high speeds is becoming increasingly critical as digital economies generate enormous data. The current paradigms for timely data processing are edge computing and data stream processing (DSP). Edge computing places resources closer to where data is generated, while stream processing analyzes the unbounded high-speed data in motion. However, edge stream processing faces rapid workload fluctuations, complicating resource provisioning. Inadequate resource allocation leads to bottlenecks, whereas excess allocation results in wastage. Existing reactive methods, such as threshold-based policies and queuing theory scale only after performance degrades, potentially violating SLAs. Although reinforcement learning (RL) offers a proactive approach through agents that learn optimal runtime adaptation policies, it requires extensive simulation. Furthermore, predictive machine learning models face online distribution and concept drift that minimize their accuracy. We propose a three-step solution to the proactive edge stream processing autoscaling problem. Firstly, a GRU neural network forecasts the upstream load using real-world and synthetic DSP datasets. Secondly, a transfer learning framework integrates the predictive model into an online stream processing system using the DTW algorithm and joint distribution adaptation to handle the disparities between offline and online domains. Finally, a horizontal autoscaling module dynamically adjusts the degree of operator parallelism, based on predicted load while considering edge resource constraints. The lightweight GRU model for load predictions recorded up to 1.3\% SMAPE value on a real-world data set. It outperformed CNN, ARIMA, and Prophet on the SMAPE and RMSE evaluation metrics, with lower training time than the computationally intensive RL models. 
\end{abstract}


\section{Introduction}
The emergence of edge computing and the Internet of Things (IoT) has resulted in unprecedented streams of data being generated at the periphery of the cloud network. Edge stream processing has become the standard for analyzing continuous data flows with minimal latency. Edge computing is an extension of the centralized cloud network for a more rapid, timely, and secure data processing \cite{cao2020overview}. A stream is an unbounded sequence of data units (tuples) $x_1, x_2, x_3, \dots$ selected from a data source. Real-time processing of this infinite stream of tuples at the resource-constrained edge presents scalability challenges. Unlike the traditional data analytics paradigm, where bounded records with known metrics are retrieved for processing, DSP handles data in motion with little foreknowledge of the incoming workload’s metrics. Manual tuning of SPS parameters such as parallelism levels, memory, CPU, timeouts, or buffer sizes for optimal application performance is time-consuming \cite{bilal2017towards} with a complexity classified as NP-hard by \cite{jonathan2020wasp} and \cite{herodotou2022automatic}. Over-allocation wastes costly cloud resources while under-allocation leads to bottlenecks, violating QoS requirements such as latency \cite{russo2023hierarchical}. Elastic scaling is required for efficient resource management that facilitates low-latency processing. The data stream is scaled along two dimensions, i.e., vertically by adjusting compute resources such as CPU or memory, and horizontally by adjusting the degree of operator parallelism.  

\begin{figure}[htbp]
    \centering
    \includegraphics[width=1.0\linewidth]{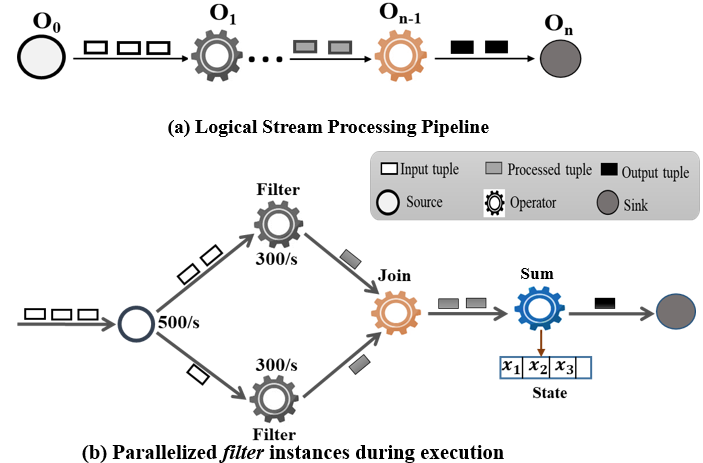}  
    \caption{Logical and Runtime Graphs of a DSP Job}
    \label{fig:datastream}
\end{figure}

This work focuses on horizontal scaling since operators constitute the core processing elements in the data stream. The dataflow topology of a DSP application is modeled as a Directed Acyclic Graph (DAG), i.e. $G_{\text{app}} = (V_{\text{ops}}, E_{\text{streams}})$, 
where $V_{\text{ops}}$ are vertices representing operators and $E_{\text{streams}}$ are directed edges representing the data flow.  Figure~\ref{fig:datastream}a,  illustrates a logical stream processing pipeline where the source operator $O_0$ ingests data, while downstream operators $O_1$ to $O_{n-1}$ process the stream. They apply processing logic e.g. user-defined functions, machine learning, and stream processing operations such as \textit{join}, \textit{map}, \textit{filter}, \textit{sum}, \textit{window}, to transform the input stream. Each operator emits a new stream of tuples as output to a downstream operator or a sink $O_n$. However, at runtime in (b), the \textit{filter} operator is replicated into two parallel instances to sustain the source operator's output rate of 500 tuples per second while the processing rate of the operator is 300 tuples per second. The \textit{sum} operator is stateful since it collects multiple tuples and updates memory state to produce results while a stateless operator like filter processes each tuple independently without memory of past tuples. Stateful operators perform tasks with high computational overheads \cite{siachamis2023towards}.
We specifically tackle the challenge of proactively deciding the optimal number of parallel operator instances to be executed in order to sustain the incoming load, as well as selecting the appropriate execution environment within the IoT-edge-cloud architecture.
Existing approaches such as Machine learning, threshold-based policies, reinforcement learning, and heuristics, are either reactive, resource-demanding, or lack runtime adaptation to the data stream. For example, threshold-based techniques monitor coarse-grained performance metrics such as CPU or memory utilization and trigger scaling based on predefined bounds. This approach is reactive due to delays in metric reflection and propagation. It can also lead to inaccurate scaling due to multi-application dependence on the same DSP computing resource under observation \cite{kalavri2018three}. Proactive machine learning approaches e.g. \cite{wu2017fas},\cite{hu2019stream}, typically do not accommodate runtime concept drifts \cite{gunasekara2024recurrent} and distribution shifts \cite{heinze2021elastic} in data streams. Reinforcement learning approaches \cite{russo2023hierarchical,xu2021model} offer an adaptation mechanism but at a high computational cost. 

An adaptive three-step framework is proposed in this work to proactively scale the data stream while optimizing the convergence rates and sample efficiency compared to existing models. The framework consists of a predictive stage, a transfer learning stage, and an autoscaling stage.  We explore a lightweight GRU neural network as the predictive model for forecasting the non-stationary ingress rates(load) along with other TSF baselines such as convolutional neural networks (CNNs), ARIMA, and Facebook’s Prophet. The models are fitted on real-world and synthetic DSP datasets at different sampling rates, i.e. 5 minutes, 2 minutes, and 1 minute. The lightweight GRU neural network achieved the best performance across datasets, with an optimal SMAPE value of 1.3\% on the 5-minute real-world dataset and an average of 3.5\% across all the datasets after 10 training epochs. An average end-to-end training and inference time of 218.8 seconds was recorded across all datasets, which is a substantial improvement on the extensive training required in reinforcement learning. We adapt the offline predictive model to an online DSP system at the edge by handling distribution disparities through a homogeneous transductive model-based transfer learning framework. The dynamic time-warping algorithm is used to identify and select aligned time series between the source and target domains. In addition, distribution shifts and concept drifts are handled by minimizing the Maximum Mean Discrepancy (MMD) and Conditional Maximum Mean Discrepancy (CMMD). Finally, the autoscaling framework calculates the minimum number of online DSP operators required to handle the forecasted ingress rates in each time step. It follows the MAPE-K loop of autonomic systems with a load balancer as the knowledge base to resolve the resource constraints at the edge.

The main contributions of this work are summarized as follows:

\begin{enumerate}
    \item We develop a lightweight GRU neural network that accurately predicts the non-stationary edge load within each tumbling time window over a projection horizon. The edge stream processing load is simulated using aggregation, interpolation, and random noise induction.
    
    \item A transfer learning framework that adapts an offline time series forecasting model to an online stream processing system. It handles the disparities in marginal and conditional distributions of the source and target domains for effective knowledge transfer.
    
    \item A hybrid proactive edge stream processing autoscaling framework that determines the optimal parallelism level for each operator in the dataflow graph. It also integrates a novel load balancer that migrates operations causing persistent backpressure (e.g., complex stateful operators) to the cloud, based on trade-offs between edge and cloud processing latency.
\end{enumerate}

The paper is subsequently structured as follows: Section II highlights relevant state-of-the-art literature on the topic. In Section III, we introduce fundamental concepts in data stream processing and provide a background to the edge autoscaling problem. The the three-step proactive autoscaling framework is discussed in Section IV with a detailed description of the various modules. The predictive module is implemented and evaluated in Section V, together with other TSF techniques as baselines. We present an analysis of the experimental results in Section VI. The paper is concluded in Section VII.

\section{RELATED WORK}
We summarily review existing works in this section and identify gaps that motivate this research. Considerable research efforts have been spent on elastic scaling using techniques such as threshold-based policies, reinforcement learning, control theory, time-series forecasting, queuing theory, and heuristic algorithms. The majority of existing works target the horizontal scaling dimension \cite{cardellini2022runtime}, although
\cite{russo2021mead,hoseinyfarahabady2020q,singh2020auto,mei2020turbine} scales DSP applications vertically. While horizontal scaling provides almost unlimited elasticity, the vertical dimension is delimited by the maximum capability of the processing node. \cite{de2020optimal,peng2019joint} combines both dimensions through operator replication and resource tuning, while \cite{liu2022elastic,cardellinioptimal}  handles the operator placement and parallelization problem simultaneously. The threshold-based techniques in \cite{gulisano2012streamcloud,gedik2013elastic,russo2021elastic,heinze2021elastic} are generally reactive. They utilize feedback control loops, where autoscaling actions are triggered in response to performance metrics exceeding or falling below predefined bounds. A symbiotic approach was proposed by \cite{lombardi2017elastic} that scales intermittently in either the operator parallelization or resource utilization domains, or jointly based on trade-offs. \cite{cardellinioptimal} uses integer linear programming and heuristics to optimize performance metrics and reconfiguration cost jointly while handling resource heterogeneity. Queuing theoretic approaches are proposed in \cite{de2017proactive,fu2017drs,lohrmann2015elastic,tolosana2016feedback,cooper2021stream,wang2017model}. The load on a DSP server is modeled as a dual queue in \cite{cooper2021stream} with a timer and batch mechanism for data transfer from the external to internal queue. \cite{fu2017drs} employs an M/M/C queue with each operator acting as a node within a Jackson open queuing network whereas \cite{lohrmann2015elastic}  represents each task in the DAG topology of the data stream as an M/M/1 queue in their latency-aware parallelization technique. A control-theoretic method is proposed in \cite{de2017proactive} to achieve elasticity and energy efficiency in multicore DSP systems based on forecasted load. The Predictive module in \cite{lombardi2017elastic} also utilizes ANN to forecast the input load. However, these approaches are not fully proactive, e.g. \cite{lombardi2017elastic} also incorporates a reactive module where metrics are collected and the current input load is used instead of the predicted load.\\ This work considers proactive elasticity in the horizontal dimension and its runtime adaptation to the fluctuating data stream.  \cite{kalavri2018three,russo2023hierarchical} scales DSP applications horizontally by computing the number of parallel operator instances to be executed in the runtime graph of the DSP Job. The former calculates the parallelism level of each operator in the DAG using the output and processing rates of connected operators, and the latter proposes a layered horizontal approach using Bayesian optimization and reinforcement learning. The Bayesian module controls the operator and application level elasticity while an actor-critic RL is used to derive the scaling decision. Though \cite{russo2023hierarchical} considers heterogeneity, it is not well-suited for the edge due to the assumption of readily accessible nodes, which is often not the case in edge stream processing. \cite{xu2021model} uses the upper confidence bound technique to handle the exploration and exploitation problem in RL for faster convergence, sample efficiency, and edge adaptation. Though the model’s convergence rate outperforms other baselines, it requires three thousand iterations to reach a satisfactory average reward. The extensive training time can be improved with lightweight predictive models and adaptation mechanisms. \cite{zacheilas2015elastic} relies on the Gaussian process to forecast the ingress rates for DSP elasticity, while \cite{hu2019stream} employs support vector regression to predict the load in each processing window. \cite{gontarska2021evaluation} evaluated the feasibility of using univariate multi-step time series forecasting methods to predict the DSP load. The authors identified deep learning TSF models as optimal for predicting distributed DSP load while highlighting their potential applications in dynamic scalability. \cite{kalim2019caladrius} employs Facebook's Prophet model to forecast future DSP loads in distributed stream processing setups for earlier detection of bottlenecks such as backpressure. \cite{farahabady2017qos} integrate the ARIMA TSF method in their prediction model and \cite{wu2017fas} utilizes least mean squares. Machine learning predictions may deviate from the actual load in the online DSP system due to the earlier discussed distribution shifts and concept drifts. \cite{wu2022cost} proposed a binary decision framework that determines whether to transfer pre-trained models to the online stream or build a new model from scratch for runtime adaptation by assessing cost-benefit tradeoffs without incorporating any adaptation mechanism. \cite{du2019multi} uses an ensemble transfer learning method that combines multiple models. A weighted majority voting allocates higher weights to ensembles that are more related to the current concepts. It assumes that the source and target domains are similar, but this assumption may not hold in edge environments with concept drift.

\section{BACKGROUND}
DSP applications are executed across diverse underlying infrastructures, ranging from multicore standalone servers to distributed environments such as cloud, fog, and edge \cite{cardellini2022runtime}. DSP applications deployed at the edge for IoT sensor data analytics are usually time-critical.   Offloading all the DSP jobs to distant cloud servers is suboptimal, as bandwidth limitations and long backhaul transmission times may lead to violation of QoS requirements. The IoT-Edge-Cloud setup introduces an intermediate layer that bridges the gap in the traditional IoT-Cloud pipeline. A typical edge cluster consists of distributed nodes ranging from micro data centers, edge servers, routers, and gateways with different capacities that provide compute support and network function virtualization. The edge environment is distributed, resource-constrained, and heterogeneous with a volatile runtime \cite{xu2021model}. The heterogeneity emanates from variability in computational power of edge devices, bandwidth variations, and architectural differences in the edge and cloud networks\cite{cheng2023proscale}. Stream processing jobs executed on these nodes are required to ingest, process, and forward data streams in real time. Edge stream processing systems often use lightweight publish-subscribe protocols such as MQTT-based message brokers, which favor low-latency processing over strict flow control. This results in unthrottled, bursty input streams directly proportional to the event generation rate of the external data sources that must be handled in real-time. The ingress rate is one of the fine-grained metrics for scaling the DSP application in the face of non-stationarity. It is measured as the total number of events arriving at the source operators of the DAG per unit time. The magnitude of the ingress rate measurements is a true univariate reflection of the actual load on the stream processing system. We quantify the ingress rate by partitioning the sequential stream data into fixed-sized intervals $(w_1, w_2, \dots, w_n)$ using a tumbling time window $w$. For each window (sampling rate) $w_i$, the observed ingress rate $\varpi_i$ forms a time series $\{\varpi_i\}_{i=1}^{n}$ that represents historical load behavior.

\subsection{The Elastic Edge Problem}
The proactive scaling problem at the edge is to forecast future ingress rates over a projection horizon $p$ using an accurate multi-step time series forecasting model that transforms the historical load into predictions $\{\hat{\varpi}_i\}_{i=n+1}^{n+p}$, and then uses these predicted values to determine optimal scaling actions.

Formally, the objective is to learn a forecasting function $f$ such that:
\[
f\left(\{\varpi_i\}_{i=1}^{n}\right) \approx \{\hat{\varpi}_i\}_{i=n+1}^{n+p}
\]

This forecasting model $f$, trained offline, must be adapted to the edge stream processing environment to enable proactive, resource-aware operator parallelization that meets QoS requirements. The horizontal autoscaling challenge involves determining the number of parallel instances of each operator required to achieve peak throughput under the predicted incoming online load $\hat{\varpi}_i$.
   
\subsection {Autoscaling Requirements}
Based on the characterization of the edge environment, a scaling framework that adjusts operator parallelism for real-time processing must satisfy the following requirements;
\begin{itemize}
\item 	Low-latency runtime adaptation: Scaling decisions should be executed in real time, as a response to performance degradation. This demands proactive autoscaling where predictive models forecast the load on the system instead of reacting to threshold breaches. Predictive models enable a preemptive degree of operator parallelism adjustments when they are combined with performance optimization policies.
\item	Dynamic Scalability: An edge stream processing autoscaling framework must exhibit elasticity. Its scale-up/down operations must autonomously adjust the logical and physical dataflow graphs of the job with minimal reconfiguration overheads. This is achievable through continuous online adaptation mechanisms with regard to the stochastic nature of edge workload conditions.
\item Resource Efficiency: edge stream processing autoscaling requires the deployment of lightweight models with faster convergence and sampling efficiency. Extracting meaningful patterns under short delays from fewer input samples minimizes CPU time, bandwidth, memory, and energy usage during online learning in the resource-constrained edge domain.
\item 	System Distribution and Heterogeneity: The IoT-Edge-Cloud setup is inherently distributed and hierarchically heterogeneous. Locally distributed nodes with bandwidth and hardware disparities form the first tier. Consequently, an edge stream processing auto-scaler must also account for the second-tier disparities in the edge-to-cloud backhaul network. 
\end{itemize}

\section{The Three-Step Framework}
\begin{figure}[htbp]
    \centering
    \includegraphics[width=0.99\linewidth]{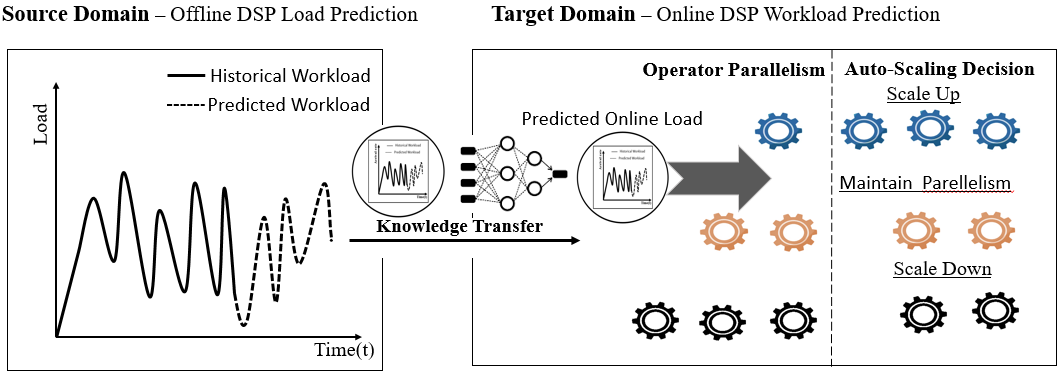}  
    \caption{Overview of the proposed approach: a hypothetical actual and predicted load, knowledge transfer, and auto-scaling}
    \label{fig:Framework}
\end{figure}

The proposed framework follows the outlined requirements to horizontally scale the edge stream processing application. Figure~\ref{fig:Framework} shows the proposed approach using hypothetical inputs and outputs. A detailed description of each step is provided in this section. While we implement and evaluate the predictive model, both the transfer learning and autoscaling frameworks are currently at advanced conceptual stages guided by theoretical principles and design considerations.

\subsection{The Proactive module}
We explore deep neural networks and conventional TSF models to predict the load on the source operators in the dataflow graph. For the deep learning TSF we review the Recurrent Neural Network (RNN) and its specialized extensions i.e. Long Short-Term Memory (LSTM) and Gated Recurrent Units (GRU) together with the Convolutional Neural Network (CNN). These neural networks have been used extensively for sequential data, whereas traditional Feedforward Neural Networks (FNNs) are limited to static data. 
\subsubsection{Deep Learning for TSF}
DSP applications run continuously for long periods, and the proactive module requires neural networks that can handle the vanishing gradient problem. The GRU and LSTM neural network extensions of the RNN effectively mitigate the long-range dependency issue in RNNs. LSTM networks integrate three gates, i.e., the forget gate $F_g$, input gate $I_g$, and output gate $O_g$, which retain essential long-term information and discard irrelevant data. However, LSTMs are computationally expensive to train due to their architectural complexity. The GRU neural network reduces training time by merging the forget and input gates into a unified \textit{update gate}~\cite{ashraf2021introduction}. The \textit{reset gate} in the GRU architecture determines how much of the previous state to forget, and the update gate determines how much new information to add to the state. GRUs maintain only a hidden state as internal memory and do not use a separate cell state $C_s$ as in LSTMs.

Given $x_t$ as input at each time step $t$ and $h_{s(t-1)}$ as the previous hidden state, the computations for the reset gate $R_g$, update gate $U_g$, candidate hidden state $\tilde{h}_s$, and final hidden state $h_s$ in a GRU are expressed as follows:

\begin{align*}
R_g &= \sigma(W_R [h_{s(t-1)}, x_t] + b_R), \\
U_g &= \sigma(W_U [h_{s(t-1)}, x_t] + b_U), \\
\tilde{h}_s &= \tanh(W_h [R_g \cdot h_{s(t-1)}, x_t] + b_h), \\
h_s &= (1 - U_g) \cdot h_{s(t-1)} + U_g \cdot \tilde{h}_s,
\end{align*}

where $W_R$, $W_U$, and $W_h$ are the weight matrices, and $b_R$, $b_U$, and $b_h$ are the corresponding bias vectors that regulate the reset gate, update gate, and candidate hidden state respectively~\cite{torres2021deep}.

\begin{figure}[htbp]
    \centering
    \includegraphics[width=0.99\linewidth]{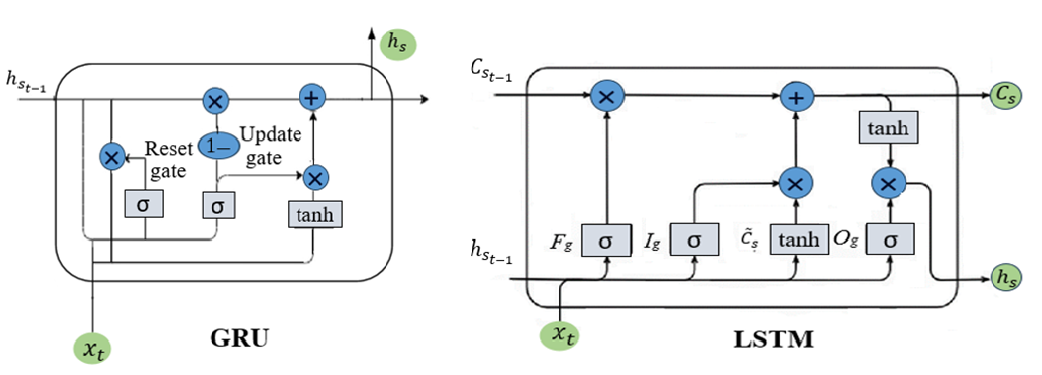}  
    \caption{ Architectures of the GRU and LSTM Neural Networks.}
    \label{fig:LSTM_GRU}
\end{figure}

LSTM and GRU architectures use the sigmoid activation function $\sigma$ to control information flow through their gates, and the $\tanh$ function to introduce non-linearity  as in Figure~\ref{fig:LSTM_GRU}.

Aside the computational efficiency of the GRU neural network, it also achieves comparable and, in most instances, better accuracy to that of the LSTM neural network \cite{mirzaei2022comparative,zarzycki2021lstm,yamak2019comparison}. We therefore settle on the GRU neural network as the optimal lightweight variant of the RNN for our ESP load predictions. 

Convolutional Neural Networks (CNNs) are also compelling architectures for time series forecasting. They can learn local patterns and features by applying convolutional filters across sequential data \cite{ye2021implementing}). CNNs can handle noise and outliers while its convolutional layers extract relevant representations from the input data. Hence, we evaluate its performance on the real-world and synthetic DSP loads alongside the GRU as deep learning paradigms.

\subsubsection{Conventional Models}

ARIMA is a linear model for time series forecasting. It combines three components: denoted as ARIMA$(p, d, q)$, where $p$ is the number of autoregressive lags, $d$ is the differencing order, and $q$ is the number of moving average lags~\cite{pandey2019predictive}. The ARIMA model can be written as a linear function of the form:

\[
Y_t = c + \phi_1 Y_{t-1} + \phi_2 Y_{t-2} + \theta e_{t-1} + e_t
\]

where $Y_t$ is the value of the time series at time $t$, $c$ is a constant, $\phi$ is the autoregressive parameter, $\theta$ is the moving average parameter, and $e_t$ is the error term at time $t$.

\vspace{1mm}
\textbf{Prophet} is based on a generalized additive model and is capable of handling missing data, outliers, and seasonality~\cite{rafferty2021forecasting}. Prophet models a time series $Y_t$ as:

\[
Y_t = G_t + S_t + H_t + \varepsilon_t
\]

where $G_t$ represents the trend, $S_t$ the seasonality, $H_t$ the holiday effects, and $\varepsilon_t$ the noise. Prophet also supports custom user-defined features.
\vspace{1mm}
\subsection{The Transfer Learning Framework}

Transfer learning (TL) simply imply the application of knowledge gained from performing one task to a different but related task. We propose a TL-based domain adaptation architecture to resolve the real-world distribution and context shifts that may affect the accuracy of the lightweight TSF model. Also, Traditional databases used for storing time series data retains the data for shorter periods e.g. 7 days for InfluxDB. This limits the availability of long-term runtime data in ESP environments. It requires the use of transfer learning to pre-train models on extensive offline dataset before adaptation to the online system.We define two domains, i.e. $\mathcal{D}_{\text{of}}$ as the source domain and $\mathcal{D}_{\text{on}}$ as the target domain. The goal is to retrieve knowledge from a related $\mathcal{D}_{\text{of}}$ and the offline task $\mathcal{T}_{\text{of}}$ to optimize an online predictive function $f_p$, while avoiding negative transfer in the target domain where $\mathcal{D}_{\text{of}} \ne \mathcal{D}_{\text{on}}$.

The transfer is considered homogeneous, because the feature and label spaces are the same: $X_{\text{of}} = X_{\text{on}}$ and $Y_{\text{of}} = Y_{\text{on}}$. However, differences may exist in the marginal and conditional distributions:
\[
P(X_{\text{of}}) \ne P(X_{\text{on}}), \quad P(Y_{\text{of}} \mid X_{\text{of}}) \ne P(Y_{\text{on}} \mid X_{\text{on}})
\]

A limited preliminary data collection phase in the target domain produces a target series $T_S$ that captures real ingress patterns. Let $S_S = \{S_{S1}, S_{S2}, \ldots, S_{Sn}\}$ be a set of candidate source time series from the historical offline datasets. A DTW distance threshold $d_t$ is defined such that: 
\[
\text{DTW}_d(S_{si}, T_S) < d_t
\]

Candidates that exceed the threshold are returned to the pool of potential datasets. This condition ensures that the selected source series for training and testing the offline model are similar enough to the target stream. However, DTW is shape-based and does not consider intrinsic distribution differences~\cite{ye2021implementing}. Relying solely on fine-tuning yields inaccurate predictions when notable distributional discrepancies exist between the source and target domains~\cite{lu2023multi}.

A 1D convolutional neural network (1D-CNN) extracts features from both source and target time series to handle the discrepancies. These features are mapped into a Reproducing Kernel Hilbert Space (RKHS) using kernel mean embedding, where the MMD and CMMD can be calculated. For example, given the extracted features be $X_{\text{of}} = \{a_i\}_{i=1}^n$ and $X_{\text{on}} = \{b_j\}_{j=1}^m$, sampled from $P_s$ and $P_t$, respectively. The maximum mean discrepancy ($\mathbb{M}$)  between $X_{\text{of}}$ and $X_{\text{on}}$ is calculated in  as :

\begin{align*}
\mathbb{M}^2 &= \frac{1}{n^2} \sum_{i=1}^n \sum_{i'=1}^n k(a_i, a_{i'}) 
+ \frac{1}{m^2} \sum_{j=1}^m \sum_{j'=1}^m k(b_j, b_{j'}) \\
&\quad - \frac{2}{nm} \sum_{i=1}^n \sum_{j=1}^m k(a_i, b_j)
\end{align*}

where $k$ is a kernel function~\cite{ouyang2021maximum}. MMD quantifies the difference in marginal distributions $\mu P_s$ and $\mu P_t$, and CMMD calculates conditional distributions, represented in the RKHS as $\mu(Y_{\text{of}} \mid X_{\text{of}})$ and $\mu(Y_{\text{on}} \mid X_{\text{on}})$.

Figure 4 illustrates the various processes in the TL framework. During joint distribution adaptation, the MMD and CMMD losses are added to the task-specific loss $L_t(\theta)$. The overall objective function $L_{\text{joint}}$ minimizes the sum of the task-specific loss $L_t(\theta)$, the MMD loss $L_{\mathbb{M}}(\theta)$, and the CMMD loss $L_{\text{CMMD}}(\theta)$ with respect to the model parameters $\theta$, each weighted by corresponding hyperparameters $\lambda_1$ and $\lambda_2$ as:

\[
\min_{\theta} L_{\text{joint}} = \min_{\theta} \left( L_t(\theta) + \lambda_1 L_{\mathbb{M}}(\theta) + \lambda_2 L_{\text{CMMD}}(\theta) \right)
\]
\begin{figure*}[t]
    \centering
    \includegraphics[width=\textwidth]{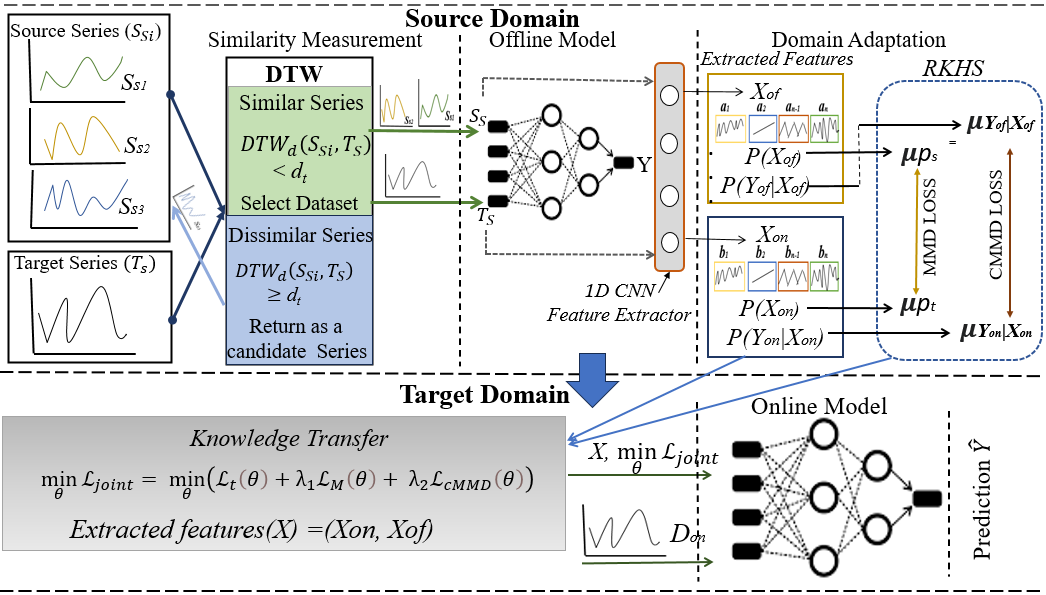}
    \caption{Process Flow of the proposed TL Framework}
    \label{fig:tl_framework}
\end{figure*}

Optimizing the combined auxiliary and task-specific losses supports effective knowledge transfer and reduces the need for frequent model updates. At minimum MMD and CMMD, the difference between the marginal and conditional distributions in the source and target domains is zero, i.e:
\[
P(X_{\text{of}}) = P(X_{\text{on}}), \quad P(Y_{\text{of}} \mid X_{\text{of}}) = P(Y_{\text{on}} \mid X_{\text{on}})
\]

The adapted architecture, weights, and learned features of the pre-trained model are retained as initialization points and fine-tuned to the current online stream for predictions in the target domain.

\subsection{The Horizontal Autoscaling Framework}
The runtime graph of a DSP job must sustain the overall ingress rate in each processing window. This is achieved by replicating operators into parallel instances such that their combined processing rate matches the input rate of each assigned task, thereby avoiding backpressure. We adopt the parallelism formulation proposed by \cite{kalavri2018three} and adapt it for the edge environment.

The true processing rate $p_{ij}$ and output rate $\sigma_{ij}$ of an operator instance $O_{ij}$ are given by:

\[
p_{ij} = \left\lceil \frac{\Gamma_p}{\tau} \right\rceil, \quad \sigma_{ij} = \left\lceil \frac{\Gamma_o}{\tau} \right\rceil
\]

where $\Gamma_p$ and $\Gamma_o$ represent the number of records processed and emitted during a time window $\tau$, which includes serialization, processing, and deserialization time.

For an operator $O_i$ with $k$ replicas, its total processing and output rates are aggregated as:

\[
p_i = \sum_{j=1}^{k} p_{ij}, \quad \sigma_i = \sum_{j=1}^{k} \sigma_{ij}
\]

A downstream operator $O_i$ receives inputs from connected upstream operators $O_j$. We define an indicator function $I(O_i, O_j)$ as:

\[
I(O_i, O_j) =
\begin{cases}
1, & \text{if } O_i \text{ and } O_j \text{ are adjacent} \\
0, & \text{otherwise}
\end{cases}
\]

The minimum degree of parallelism $\eta_{oi}$ for each operator is calculated as:

\[
\eta_{oi} = \left\lceil \frac{\sum_{j=1}^{i-1} I(O_i, O_j) \cdot \sigma_j}{p_i / k} \right\rceil, \quad i < n
\]
\begin{figure*}[t]
    \centering
    \includegraphics[width=\textwidth]{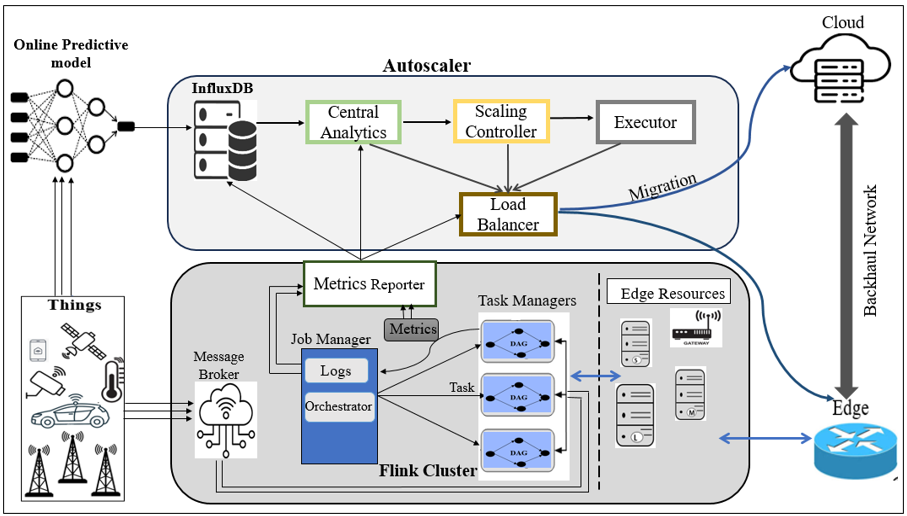}
    \caption{Architecture of the Proposed Autoscaler}
    \label{fig:tl_framework}
\end{figure*}

This ensures that downstream operators are provisioned to handle the aggregated output rate of their upstream dependencies.However, this concept has limitations in DSP systems. Increasing parallelism does not always result in higher throughput due to inter-operator resource competition and runtime overheads. For instance, excessive parallelism reduces per-task memory, which can increase disk or network I/O frequency, especially for stateful operators~\cite{han2025learning}. In such cases, built-in flow control mechanisms such as backpressure may throttle data flow, introducing additional latency even after scaling. Complex or stateful operations are generally not ideal for edge-only execution due to the limited computational capacity of edge devices~\cite{silva2019investigating}.

To adapt this mechanism for edge environments, we define thresholds for a load balancer which determines if an operator has reached its maximum parallelism limit $\eta_{oi}^{\text{max}}$. For a given task execution, latency tradeoffs are considered such that if the edge latency $\varepsilon_l$ exceeds the combined migration and cloud latency, i.e.,$\varepsilon_l > M_l + C_l$
then $\eta_{oi} = \eta_{oi}^{\text{max}}$, and the task is offloaded to the centralized cloud as illustrated in Figure 5. This approach balances resource constraints at the edge with latency QoS requirements of the overall DSP system.
\subsubsection{The Mape-K loop Autoscaling Architecture}
The autoscaling framework follows the MAPE-K loop from autonomic computing, which consists of the phases: Monitor, Analyze, Plan, Execute, and Knowledge ~\cite{dehraj2021review}. These components of the autoscaler are described using the Apache Flink stream processing engine as a use case:
\paragraph{Monitor} In the initial phase, performance metrics of the streaming job are collected. In Flink, Task Managers report task status, metrics, and statistics to the Job Manager. The metrics reporter retrieves the latest indicators from the Task Managers based on the engine's instrumentation. These metrics, along with logs from the Job Manager and the forecasting model's predictions, are stored in the InfluxDB time series database. The predicted rates and collected metrics form the basis for the next phase.
\paragraph{Analyze} The analytics module runs advanced queries on the stored metrics to extract actionable insights. Key indicators include throughput, processing latency, and backpressure. This phase evaluates the current system configuration to support scaling decisions.

\paragraph{Plan} Based on the analysis output, the scaling controller formulates an autoscaling policy such as scaling up, scaling down, or maintaining the current state. When an adjustment is required, the optimal degree of operator parallelism $\eta_i$ is computed to adapt to workload changes while optimizing resource usage.

\paragraph{Execute} If the scaling controller recommends adjustments, the running job is temporarily halted and checkpointed for fault tolerance and to minimize reconfiguration overheads. The job is then restarted with the updated parallelism configuration, applying the autoscaling decision with minimal interruption to the stream processing pipeline.

\paragraph{Knowledge} The load balancer acts as the knowledge repository of the MAPE-K loop. It collects real-time updates from all components, including the metrics reporter, and manages edge resources effectively. When an edge resource reaches its limit (i.e., $\eta_{oi} = \eta_{oi}^{\text{max}}$), the load balancer initiates migration of the streaming job from the edge operator to the cloud.

\vspace{1ex}
The autoscaling framework operates in parallel with the host stream processing engine, without interrupting the job's runtime. It only modifies the parallelism configuration during scale operations. The main inputs to the autoscaler come from the predictive model and collected metrics via InfluxDB, while the stream processing engine receives input from data sources or message brokers. Although the load balancer interacts with the system only during extreme workload conditions, its interaction occurs at a lower level of abstraction via the edge gateway. Stream processing engines are normally installed on the edge gateways that possess more computational power in the edge environment compared to the sensor networks \cite{shahid2023some}.

\section{Implementation of the Proactive Module}
Real-world and synthetic datasets were used in the experimental setup.  The synthetic IoT Traffic dataset sourced from \cite{gontarska2021evaluation} consists of simulated vehicular traffic recorded per second over 5-minute, 15-minute, and 1-hour granularities.  We also used the real-word dataset from the New York City Taxi and Limousine Commission-TLC Trip Record Data. This dataset contains millions of taxi trip records collected monthly, with each entry consisting of pickup and drop-off timestamps, coordinates, trip distances, number of passengers, and fare-related attributes. The New York City Taxi Trips (NYCTT) data has been used extensively in DSP and smart city research including \cite{xu2021model,tschumperlin2018using,agrawal2018efficient,arkian2021model}. Each of the datasets is curated and adjusted to fit the specific requirements of edge stream processing and its load projection horizons. 

\subsection{Data Preprocessing}
The original 5-minute traffic data is randomized and resampled to finer-grained temporal resolutions of 1-minute, 2-minute, and 5-minute intervals to simulate realistic loads in DSP applications at the edge through upsampling, interpolation, and probabilistic noise injection detailed in Algorithm 1. This process allows the model to adapt responsively to sudden changes in workload patterns and better capture short-term trends. A third-order spline interpolation \textit{S} that captures complex non-linear trends is used to handle missing values and to avoid overfitting. An adjustment factor $\alpha$ is used to ensure the magnitude of the loads in each sampling rate varies. Randomness is further induced in the data to replicate the stochastic nature of the edge stream using a uniform distribution. 
\begin{algorithm}[htbp]
\caption{ESP Load Simulation (IoT Traffic)}
\begin{algorithmic}[1]
\renewcommand{\algorithmicrequire}{\textbf{Input:}}
 \renewcommand{\algorithmicensure}{\textbf{Output:}}
\REQUIRE $\mathcal{D}$ : Time series data $\{t_i, \varpi_i\}_{i=1}^n$\\
\hspace*{1.3em} $\Upsilon$ : Sampling rate
\ENSURE $\mathcal{D'}:$ \text{ESP time series load}
\STATE $\mathcal{D}_{\text{res}} \gets$ Partition $\mathcal{D}$ into empty bins using $\Upsilon$
 \STATE $\tau \gets \{t_i - t_1 \mid \forall t_i \in \mathcal{D}\}$ \text{ in seconds}
\STATE  $\tau' \gets \text{New time axis } \{t_j - t_1 \mid \forall t_i \in \mathcal{D}_{\text{res}}\} \text{ in seconds}$
    \FOR{each missing $\varpi_i \in \mathcal{D}_{\text{res}}$}
        \STATE $\mathcal{S} \gets$ Fit cubic spline over $\tau$ and $\varpi_i$
        \STATE $\varpi' \gets \text{Interpolate load on new time axis } \mathcal{S}(\tau')$
    \ENDFOR

\STATE  $\alpha \gets \text{Calculate adjustment factor based on } \Upsilon$
\STATE  $\varpi' \gets \text{Adjust load values using } \alpha \cdot \varpi_j'$
\STATE  $\varpi' \gets \text{Add } \pm 10\% \text{ random noise to } \varpi'$
\STATE  $ \text{Convert } \varpi' \text{ to float64 and replace NaN}$
 \STATE  $\text{Convert } \varpi' \text{ to int}$
 \STATE Update $\mathcal{D}_{\text{res}}$ with $\tau'$, $\varpi'$
  \STATE  $\mathcal{D'}\gets \text{Reset index of } \mathcal{D}_\text{res}$
\RETURN  $\mathcal{D'}$
\end{algorithmic}
\end{algorithm}

The mean and standard deviation of the load at each resampled rate are calculated. They serve as benchmarks for shifting and scaling the real-world (NYCTT) data for comparability. 
 For the NYCTT data, four files are made available by the various providers for each month: the Yellow Taxi, Green Taxi, For-Hire Vehicle (FHV), and High Volume For-Hire Vehicle (HVFHV) Trip Records. The pickup times for all providers for January 2024 were extracted from the four parquet files and combined into one CSV file with over 4 million records.  The extracted records are further sampled into coarser granularities, detailed in algorithm 2, using the earlier defined time intervals to produce three samples. The means and standard deviations of each sample are shifted and scaled to match those of the corresponding IoT traffic dataset. 
\begin{algorithm}[htbp]
\caption{ESP Load Simulation NYCTT}
\begin{algorithmic}[1]
\renewcommand{\algorithmicrequire}{\textbf{Input:}}
\renewcommand{\algorithmicensure}{\textbf{Output:}}

\REQUIRE $\mathcal{T}: \text{ Extracted timestamps } \{t_i\}_{i=1}^n$ \\
\hspace{1.5em} $\Upsilon: \text{ Sampling rate}$ \\
\hspace{1.5em} $\mu_t, \sigma_t: \text{ Target mean and standard deviation}$

\ENSURE $\mathcal{D'}: \text{ ESP time series load}$

\STATE $t_{\text{index}} \gets \text{Set temporary index for } \mathcal{T}$

\STATE $\nu \gets \text{Partition } \mathcal{T} \text{ into uniform intervals using } \Upsilon$

\STATE $\varpi \gets \text{Count events per } \nu$

\STATE $\mathcal{D}_{\text{res}} \gets (\nu, \varpi)$

\STATE $\text{Reset index of } \mathcal{T}$

\FOR{each $\varpi_i \in \mathcal{D}_{\text{res}}$}
    \STATE $z_i \gets \text{Standardize } \varpi_i \text{ to z-scores}$
    \STATE $\varpi'_i \gets z_i \cdot \sigma_t + \mu_t$
\ENDFOR

\STATE $\mathcal{D'} \gets (\nu, \varpi')$

\RETURN $\mathcal{D'}$

\end{algorithmic}
\end{algorithm}

\subsection{Experimental Setup}
Sampling rates of 1 minute, 2 minutes, and 5 minutes are passed as arguments to each of the DSP load simulation functions in Algorithm~1 and Algorithm~2. This produces six datasets at the corresponding new sampling rates. These datasets are then used to evaluate the predictive performance and computational efficiency of the proposed GRU model, alongside other TSF methods.

Table~1 shows the runtime of the experiments. All the datasets are partitioned into 80\% for training and 20\% for testing, and this split is applied consistently across all 24 experiments conducted.

\begin{table}[ht]
\small
\centering
\caption{Configuration} 
\renewcommand{\arraystretch}{1.5}
\begin{tabular}{|l|p{5.8cm}|}
\hline
\textbf{Resource Type} & \textbf{Specifications} \\
\hline
CPU & Intel(R) Core(TM) i5-6500 CPU @ 3.20GHz, 3.19 GHz\\
\hline
RAM & 4.00 GB Physical Memory\\	
\hline
vGPU & Tesla T4 GPU (16 GB virtual Memory)\\
\hline
Software & Google Colab runtime (Python 3.11.12), PyTorch, CUDA 12.2, Scikit-learn 1.6.1, pmdarima 2.0.4 \\
\hline
\end{tabular}
\end{table}

The architecture used in the deep learning experiments for both the GRU and CNN models consists of an input sequence length of 24 time-steps, with the data normalized to a range of [0, 1]. Each model includes an input layer, two hidden layers, and a dropout layer with a dropout probability of 0.2 to mitigate overfitting. A fully connected output layer with a single unit is used to predict the load for the next time step. The models were compiled using the Adam optimizer with a learning rate of 0.01 and were trained for up to 10 epochs using a batch size of 16, minimizing the Mean Squared Error (MSE) loss function. The \texttt{zero\_grad()} method resets gradients before each batch, optimizing memory usage and ensuring correct backpropagation. The \texttt{squeeze()} function is used to align predicted and actual tensor dimensions for correct loss computation. An inverse scaling function transforms the normalized predictions back to their original scale. Model hyperparameters, such as the number of layers, number of units, learning rate, dropout rate, and sequence length, are empirically tuned through iterative experimentation.

The GRU model utilizes built-in sigmoid and tanh activation functions to capture long-term dependencies in sequential data. Each of its two hidden layers contains 64 units.

The CNN model consists of two Conv1D blocks, each consisting of a convolutional layer, a ReLU activation function, a pooling layer, and a dropout layer. The convolutional layers apply 64 kernels of size 3 to detect features such as bursts or trends. ReLU introduces non-linearity to model complex temporal patterns. The pooling layer downsamples the input by a factor of 2, followed by dropout. A flattening layer then reshapes the convolutional output into a 1D tensor, which is passed to the fully connected output layer.

For the conventional TSF models, we prepare the Facebook Prophet environment  by installing \texttt{PyStan} before installing the Prophet library to ensure compatibility. A Prophet object is created, fitted to the training segment of the time series data, and evaluated on the test portion. This fitting process involves capturing trend, seasonality, and holiday effects if applicable. The \texttt{yhat} values, representing the most likely predictions, are extracted from the model's multi-attribute output. Prophet automates the hyperparameter selection and tuning process.

The ARIMA model is implemented using the \texttt{pmdarima} library, which simplifies model selection through the \texttt{auto\_arima} function. The search space for optimal $(p, d, q)$ parameters is constrained to a maximum of 3 for $p$ and $q$, and $d \in \{0,1\}$, to reduce computational cost given the data complexity. The Nelder-Mead method is used to optimize model coefficients, with the maximum number of iterations set to 30. For each iteration, the Akaike Information Criterion (AIC) is computed, and the configuration yielding the lowest AIC is selected as the final model. The ARIMA model is further adapted to temporal changes by using a rolling window approach that updates model parameters with recent observations.

\subsection*{Evaluation}
We analyze the predictive performance of various time series forecasting (TSF) models on the test set of the simulated stream processing loads across different sampling rates. Additionally, we assess the resource efficiency of each model by measuring its end-to-end training and inference latencies. 

Our evaluation employs Symmetric Mean Absolute Percentage Error (SMAPE) and Root Mean Square Error (RMSE) as benchmark accuracy metrics. Let $n$ denote the total number of tuples in the dataset, and let $\hat{\varpi}_i$ represent the predicted load at time step $i$. The selected evaluation metrics are defined as follows:

\textbf{Symmetric Mean Absolute Percentage Error (SMAPE):} This metric is widely used for predictive tasks involving heterogeneous value scales. It expresses the error as a percentage of the actual values, making it easier to interpret and compare across data sources and sampling rates. Given $\varpi_i$ as the actual load and $\hat{\varpi}_i$ as the predicted load;

\[
\text{SMAPE} = \frac{100\%}{n} \sum_{i=1}^{n} \frac{|\varpi_i - \hat{\varpi}_i|}{|\varpi_i| + |\hat{\varpi}_i|}
\]

\textbf{Root Mean Square Error (RMSE):} RMSE captures the standard deviation of the prediction errors. It provides an absolute error measure and is particularly suitable for datasets of the same sampling rate. In our experiments, the NYCTT data is standardized using z-score normalization to match the scale of the IoT Traffic data.

\[
\text{RMSE} = \sqrt{\frac{1}{n} \sum_{i=1}^{n} (\varpi_i - \hat{\varpi}_i)^2}
\]

\textbf{Model Computation Time:} We also evaluate the computational cost of each model, measured as the total time taken to execute all operations involved in the model's training and evaluation process. This includes the forward pass, loss computation, backpropagation, optimizer steps, epoch processing, and any additional system overheads. The initial timestamp $t_0$ is recorded prior to model training, and the final timestamp $t_1$ is taken after training and inference. The total elapsed time is computed as: $\Delta t = t_1 - t_0$.

\section{Experiment Results}

\begin{figure*}[htbp]
    \centering
    \begin{subfigure}{\textwidth}
        \centering
        \includegraphics[width=\textwidth]{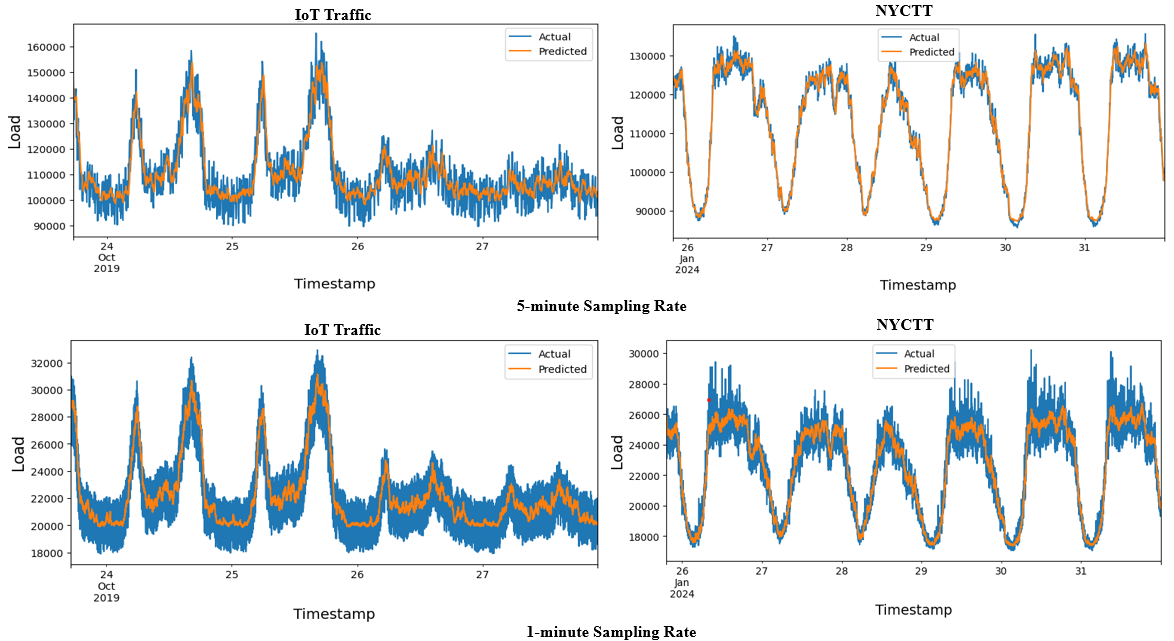}
        \caption{Forecast accuracy graph of the GRU Model}
        \label{fig:GRU}
    \end{subfigure}
    
    \vspace{1.5em} 
    
    \begin{subfigure}{\textwidth}
        \centering
        \includegraphics[width=\textwidth]{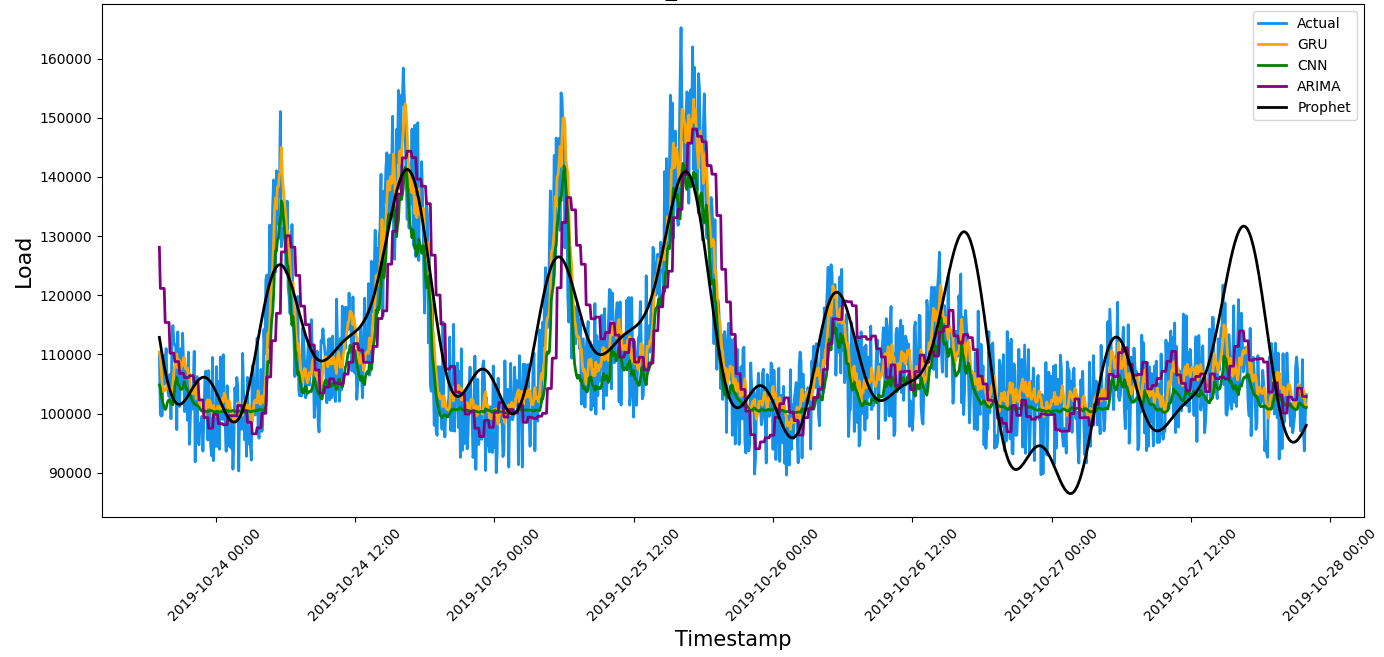}
        \caption{Forecast accuracy graph of all modesl on the 5-min IoT data}
        \label{fig:comparative}
    \end{subfigure}
    \caption{Performance comparison of GRU model and baseline methods using their forecast trajectories}
    \label{fig:gru_comparison}
\end{figure*}

The predictive performance of the lightweight proactive GRU module is first visualized by plotting its load forecast trajectories against the actual load.  Figure~\ref{fig:GRU} illustrates the model’s accuracy across varying sampling rates (1\,min and 5\,min) for both the IoT Traffic and NYCTT test sets.  Only a minimal divergence is observed between the measured and predicted loads, even in the presence of noise‑induced bursts.

\begin{figure*}[htbp]
    \centering
    \begin{subfigure}{0.49\textwidth}
        \centering
        \includegraphics[width=\linewidth]{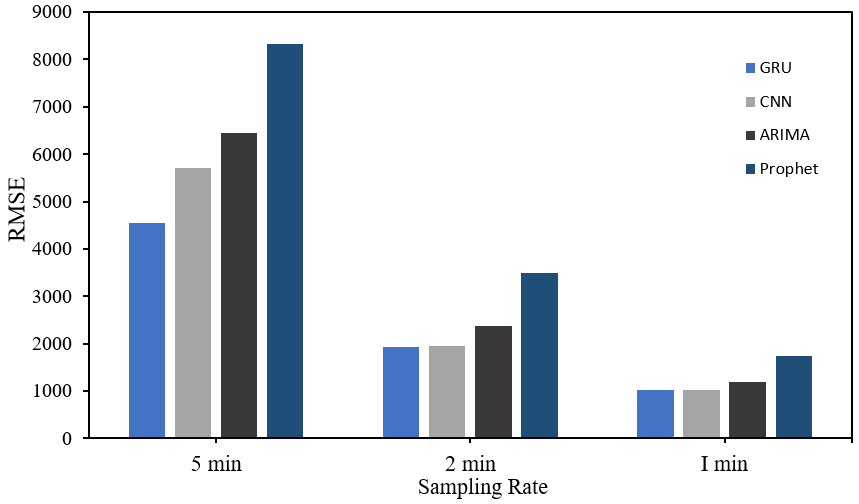}
        \caption{Average RMSE of each model across all sampling rates}
        \label{fig:RMSE}
    \end{subfigure}
    \hfill
    \begin{subfigure}{0.49\textwidth}
        \centering
        \includegraphics[width=\linewidth]{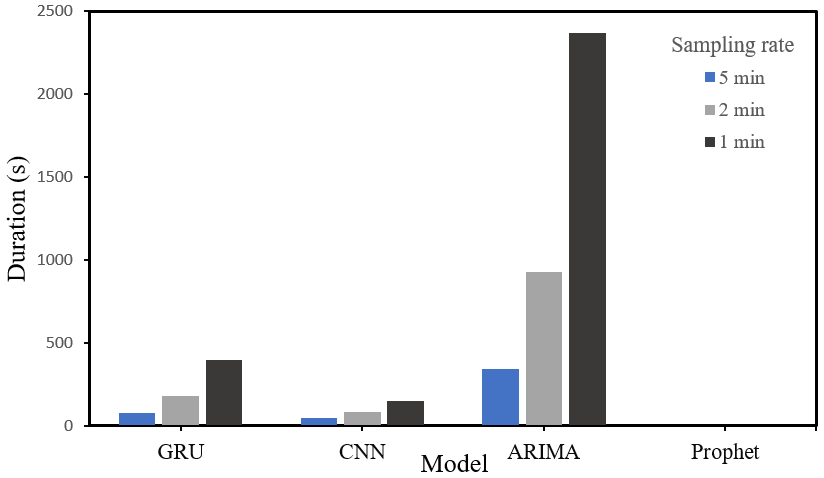}
        \caption{Average model training time for each sampling rate }
        \label{fig:training}
    \end{subfigure}
    \caption{Average RMSE and Model Compuation time of all TSF models across all datasets}
    \label{fig:combined_figures}
\end{figure*}

Figure~\ref{fig:comparative} benchmarks the GRU trajectory against the three baseline TSF methods, i.e., CNN, ARIMA, and Prophet, on the coarse‑grained 5‑minute IoT Traffic test data.  The GRU curve remains closest to the actual series, particularly during highly volatile peak periods.  In contrast, the Prophet model performs the worst among all the TSF methods. This is due to its focus on modeling long-range seasonal patterns, such as daily or hourly data, rendering it ineffective on fine-grained sampling rates with no clearly defined seasonality.

We restricted ARIMA’s search space (\(p,q\le 3,\;d\in\{0,1\}\)) and update intervals to make its training feasible in the time-constrained experimental environment. Also, the limitations ensure comparability with the other lightweight neural and conventional models. The applied constraints, however, capped its accuracy.  A broader hyperparameter search or finer parameter updates could improve ARIMA’s forecasts, though at a substantial computational cost. 

Table~\ref{tab:smape_results} reports SMAPE values and complements the visual comparisons with a quantitative measure.  Across all 24 experiments, the GRU consistently records the lowest SMAPE (highlighted in gray) except for the 2‑minute NYCTT test set, where the CNN attains marginally better accuracy.  Conversely, at the 1‑minute NYCTT rate, ARIMA slightly outperforms the CNN. All four models achieve lower errors on the NYCTT data than on the synthetic IoT Traffic data, likely because the synthetic DSP load was generated via interpolation with added noise, whereas the real‑world DSP load was processed using the original non-stationarity without additional noise induction. The best-performing model for each sampling rate is highlighted in gray, with the exceptions highlighted in yellow.

\renewcommand{\arraystretch}{1.5}
\definecolor{highlightgray}{gray}{0.85}
\definecolor{highlightyellow}{RGB}{255, 255, 153}

\begin{table}[ht]
\centering
\caption{SMAPE Evaluation Results (\%)}
\label{tab:smape_results}
\begin{tabular}{|l|l|c|c|c|c|}
\hline
\textbf{Dataset} & \textbf{Sampling Rate} & \textbf{GRU} & \textbf{CNN} & \textbf{ARIMA} & \textbf{Prophet} \\
\hline
\multirow{3}{*}{IoT Traffic} 
& 5 Min.   & \cellcolor{highlightgray}5.37 & 5.64 & 6.79 & 7.07 \\
& 2 Min.   & \cellcolor{highlightgray}5.24 & 5.35 & 6.00 & 7.14 \\
& 1 Min.   & \cellcolor{highlightgray}5.10 & 5.13 & 5.67 & 7.06 \\
\hline
\multirow{3}{*}{NYCTT} 
& 5 Min.   & \cellcolor{highlightgray}1.34 & 2.26 & \cellcolor{highlightyellow}2.19 & 4.85 \\
& 2 Min.   & \cellcolor{highlightyellow}1.82 & \cellcolor{highlightgray}1.81 & 2.40 & 5.39 \\
& 1 Min.   & \cellcolor{highlightgray}2.27 & 2.57 & 2.74 & 5.19 \\
\hline
\end{tabular}
\end{table}

 The RMSE metric is comparable between the IoT Traffic and NYCTT datasets for the same sampling rate (e.g., both 5-minute datasets), as the match statistics procedure in Algorithm~2 aligns their value distributions. However, RMSE values are not directly comparable across different sampling rates due to the scale adjustments. To address this, we normalize the RMSE values using the average absolute errors for each sampling rate. This allows for a fair comparison across the three granularities. As illustrated in Figure~\ref{fig:RMSE}, the GRU model achieves the lowest average RMSE across all sampling rates compared to the baseline methods, further confirming its superior predictive accuracy. All four models achieve lower errors on the NYCTT data than on the synthetic IoT Traffic data attributable to the data characteristics. The synthetic data was generated via interpolation with added noise, whereas the real‑world series was processed with the original non-stationarity without additional noise induction.

Figure~\ref{fig:training} presents the average model training time for each dataset across all sampling rates. The Prophet model exhibits negligible training durations due to its architecture, which does not require iterative optimization like the neural models. The ARIMA model is comparable to the GRU and CNN in terms of the training durations since it performs parameter optimization during model fitting. Despite adjusting ARIMA’s search space and update intervals to reduce computational overhead, its training time remains higher than both the GRU and CNN models across all sampling rates. The CNN model demonstrates the fastest training times overall, followed by GRU. Across all models, training time increases substantially at shorter sampling rates due to the higher volume of data processed.

\section{Conclusion}

This paper presents a proactive edge stream processing autoscaling framework designed to facilitate real-time processing in the IoT-Edge-Cloud Setup.  The proposed architecture integrates three core components: a proactive forecasting module, a transfer learning mechanism, and an autoscaling strategy. The GRU-based proactive module forecasts future loads in the data stream to inform autoscaling decision.  The predictive model is integrated into a running data stream processing system using the transfer learning framework based on joint distribution adaptation. Finally, the autoscaler adjust the operators horizontally based on a predicted workload threshold. A load balancer migrates the stream processing job from the edge to the cloud when extreme workload conditions causes backpressure. The experimental results shows that the GRU model outperforms other established TSF methods such CNN, ARIMA, and Prophet in terms of predictive accuracy. It is also computationally efficient, requiring minimal training times which is a significant improvement over the resource-intensive reinforcement learning based solutions. However, the transfer learning and autoscaling components are at the conceptual stage. Future work will explore real-world deployment, with a particular focus on inter-operator resource contention and its impact on parallelism decisions. 

\bibliographystyle{IEEEtran}
\bibliography{References}

\end{document}